\documentclass[%
reprint,
superscriptaddress,
ymb,
prl,
twocolumn,
titlepage,
floatfix,
showpacs,
APL]{revtex4-1}

\usepackage{graphicx}
\usepackage{hyperref}
\usepackage{amssymb}
\usepackage{amsmath}
\usepackage{textcomp}
\usepackage{xcolor}
\usepackage{tikz}
\allowdisplaybreaks

\definecolor{linkblue}{RGB}{49,49,148}%prl color
\hypersetup{linkcolor  = linkblue,citecolor  = linkblue,urlcolor   = linkblue,colorlinks = true,}

\makeatletter
\renewcommand*{\eqref}[1]{%
  \hyperref[{#1}]{\textup{\tagform@{\ref*{#1}}}}%
}
\makeatother

\begin{document}

\title{Rapid heat assisted polarization reversal in ferroelectric thin films}

\author{Rekikua Alemayehu}
\affiliation{Institut f\"ur Physik und Astronomie, Universit\"at Potsdam, 14476 Potsdam, Germany}
\author{Steffen Zeuschner}
\affiliation{Institut f\"ur Physik und Astronomie, Universit\"at Potsdam, 14476 Potsdam, Germany}
\author{Alexander von Reppert}
\affiliation{Institut f\"ur Physik und Astronomie, Universit\"at Potsdam, 14476 Potsdam, Germany}
\author{Matthias Roessle}
\affiliation{Helmholtz-Zentrum Berlin f\"ur Materialien und Energie GmbH, Wilhelm-Conrad-R\"ontgen Campus, BESSY II, 12489 Berlin, Germany}
\author{Marin Alexe}
\affiliation{Institute of Physics, Warwick University, UK}

\author{Matias~Bargheer}
\affiliation{Institut f\"ur Physik und Astronomie, Universit\"at Potsdam, 14476 Potsdam, Germany}
\affiliation{Helmholtz-Zentrum Berlin f\"ur Materialien und Energie GmbH, Wilhelm-Conrad-R\"ontgen Campus, BESSY II, 12489 Berlin, Germany}

\date{\today}
\begin{abstract}
We demonstrate that switching of ferroelectric thin-films sandwiched between metallic electrodes can be controlled by laser-assisted heating, reminiscent of heat-assisted magnetic recording. We employ electrical switching cycles that quantify the electrically switchable remanent polarization $P_\mathrm{r}$ and show that 300\,ns voltage pulses alone change the polarization by less than $\Delta P<P_\mathrm{r}$. Transient heating of the metallic top electrode by synchronized ns laser-pulses induces a reversal $\Delta P^\mathrm{L}>P_\mathrm{r}$ of the average polarization. The transient average temperature modeled by the heat equation can rationalize the polarization change observed for different relative timing $\Delta t$ of the laser pulse, if it arrives before the electrical pulse.

\end{abstract} \maketitle
%\section{Introduction}
Recent advances in the field of heat-assisted magnetization recording (HAMR)\cite{kryd2008} have underscored the pivotal role of thermal energy in overcoming energy barriers in ferromagnetic systems. This energy can be provided in tiny amounts via nanophotonics.\cite{chal2009} HAMR has spurred significant progress in data storage technologies - with mass production on the market since 2024 -  while providing deeper insights into the fundamental physics of ferroic materials, i.e. systems that can be switched between two states. The spontaneous polarization in ferroelectric materials can equally well represent bits, and the double-well potential landscapes exhibit striking parallels to ferromagnets, both in their functional characteristics and in the underlying mechanisms governing their behavior \cite{guo2021review}, including ultrafast photoferroic recording.\cite{kime2020} 
In particular, ferroelectric thin films have garnered widespread attention for their enhanced polarization stability, scalability, and tunability at the nanoscale.\cite{dawb2005,eder2005,schl2007} They are ideal candidates for applications in non-volatile memory devices, sensors, and energy-efficient electronic components.\cite{sett2006}  
Sufficiently large electric fields can overcome the energy barriers separating the stable polarization states. However, achieving efficient and rapid polarization switching,  remains challenging. \cite{chandra2002, dawber2003}
Light control of rapid ferroelectric switching was achieved in various ways: Photoconductive switches rapidly provide voltage.\cite{li2004} Direct photoexcitation reduces the barrier by changing the electronic structure \cite{lian2019, chen2024}, screening the depolarization field \cite{rana2009} or via interface effects.\cite{li2018} THz light fields couple directly or indirectly to the ferroelectric soft mode\cite{mank2017, chen2022} and enable switching. %This can be enhanced by using THz frequencies where the dielectric function is near zero.\cite{kwaaitaal2024epsilon} 
UV light and femtosecond lasers have been successfully applied for permanent laser poling of ferroelectrics.\cite{eggert2006,sheng2024,sarott2024,yang2018light} %A recent DFT study highlights the microscopic origin of such polarization reversal with above-bandgap excitation.\cite{Chen_DFT2024}
However, infrared pulses that heat the ferroelectric via the adjacent metal electrode adresses different physics without interband excitation of the ferroelectric.
%Such laser heating is equivalent to heating the film near to the Curie temperature, lowering the activation energy of the switching process as already shown in Merz' seminal work on BaTiO$_3$  and more recently in ferroelectric superlattices.

Indirect laser heating brings the film close to its Curie temperature, effectively lowering the activation energy for polarization switching, as initially demonstrated by Merz in seminal studies on BaTiO$_3$ \cite{Merz1954,Fatuzzo1959}, and more recently confirmed in ferroelectric superlattices \cite{sidorkin2019}.

%This can now be combined with nanoscale size and rapid transient heating:
%Developing the concept of heat-assisted polarization reversal (HAPR) in ferroelectric thin films in close analogy to HAMR offers a promising new avenue for both fundamental research and technological innovation. 

\begin{figure*}[thb!]
\centering
\includegraphics[width = 1.7 \columnwidth]{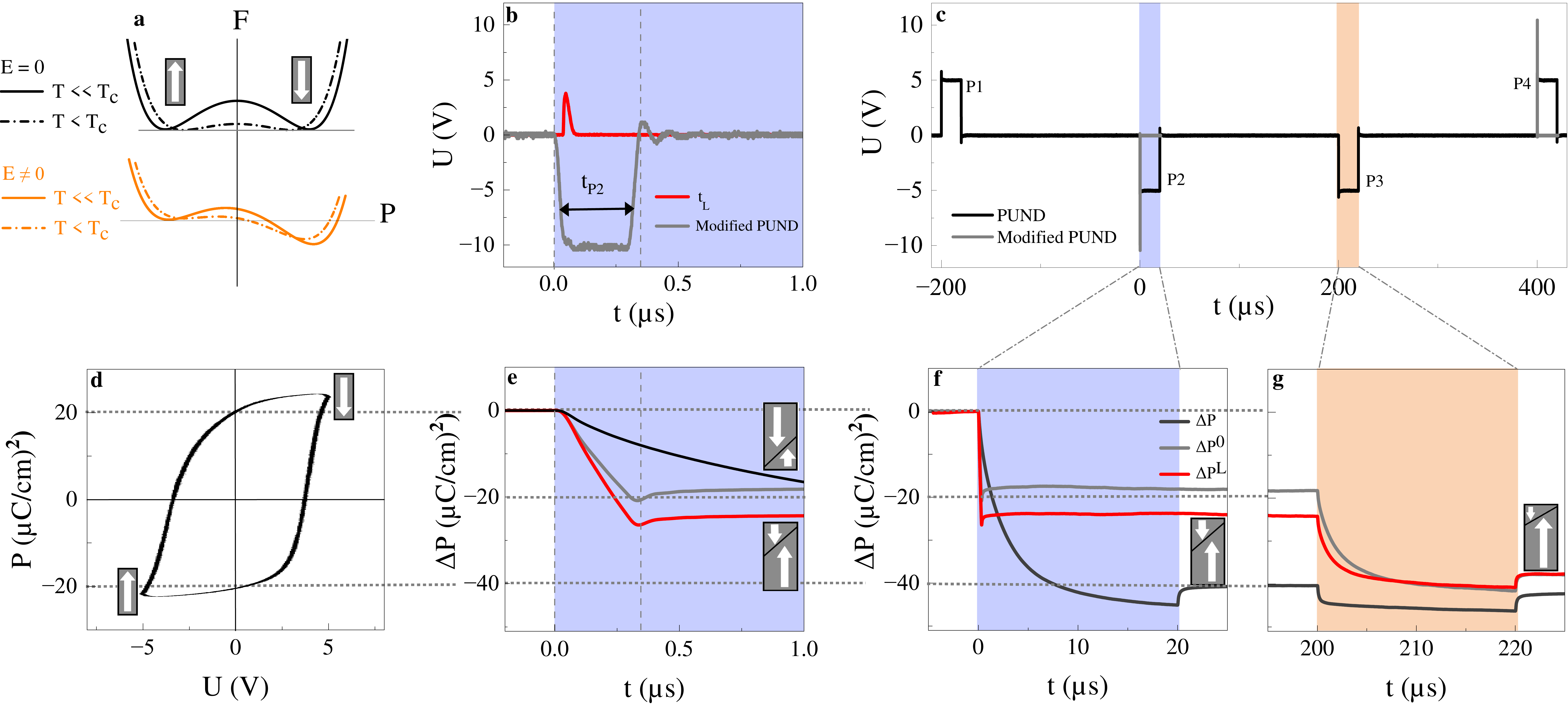}
\caption{(a) Model of the Free energy $F$ of a ferroelectric system as a function of polarization $P$ showing the effect of temperature and electric field. (b) Measured timing of short voltage pulse (grey) and synchronized laser pulse (red). (c) Pulse sequence for standard PUND (black) and modified pulse sequence with short pulses (grey) used in the laser-assisted experiments. (d) Hysteresis loop measured for a triangular waveform at 1 kHz. (e) Polarization change $\Delta P$ for 300 ns electrical pulses without (grey) and with (red) laser pulse plotted together with the polarization $P$ (black) derived from the PUND sequence shown in panel c. Note the lines indicating the timing in panels (b) and (e) and the remanent polarization $P_r$ measured in panel (d). (f) Same as (e) on a longer time scale. (g) Cross-check of the remaining switcheable polarization after the short voltage pulse with (red) and without (grey) laser. Indeed, less polarization can be switched by the long saturating pulse, if the laser had assisted the switching in the previous short pulse.}
\label{fig:Preversal}
\end{figure*}

In this study, short infrared laser pulses that do not photoexcite the ferroelectric deposit thermal energy in the metal electrode, which transfers heat to the nanoscale ferroelectric thin film, enabling rapid control of the system’s free energy landscape by light pulses. We explore heat-assisted polarization reversal (HAPR) by integrating pulsed laser excitation into a ferroelectric testing environment and show that the simultaneous excitation by thermal energy and external fields exploits the fundamental nonlinear physics associated with double-well potentials, as the result is not a superposition of the individual excitations. We find experimental conditions, where the laser-pulse alone only  produces a very weak pyroelectric response, and the electric field alone is insufficient to switch the average polarization. Only when the laser heats the film during the application of the electric field, the polarization is reversed. If the laser pulse arrives before electrical field pulse, the polarization response to both excitations depends approximately on their relative timing as predicted by one-dimensional heat conduction. Since the photon energy is below the bandgap, the laser is exclusively absorbed in the metal electrode. Thus we conclude that the energy barrier separating the two polarization states is transiently reduced by the laser-heating transferred from the electrode, facilitating polarization switching under an external electric field. This laser-induced thermal activation represents a powerful tool for manipulating the polarization dynamics of ferroelectric films and opens new doors for practical applications in energy-efficient electronics and beyond.

% \begin{figure}[thb]
%\centering
%\includegraphics[width = 0.8 \columnwidth]{fig_2.pdf}
%\caption{Setup: Ferroelectric thin film device 60\,nm PZT with 340\,$\mu$m diameter Pt top electrode is contacted via a thin tungsten (W) needle. Voltage pulse sequences generated in the function generator are synchronized to the 1\,kHz Ti:Sapphire laser system via computer-controlled delay generator. The photodiode detects the timing of the laser pulses via the light scattered from the sample. The applied voltage and the current through the resistor are measured as a voltage drop by an oscilloscope with 50 $\Omega$ impedance. The inset shows an image of the sample and the tungsten needle in contact. Optical components, including a wave plate (WP), polarizer (P), lens (L), and mirror (M), are used to direct and focus the laser beam onto the sample.}
%\label{fig:Setup}
%\end{figure}

%Figure~\ref{fig:Setup} shows the electrical circuitry used to apply voltage pulses to the thin film capacitor devices prepared by pulsed laser deposition. 
The devices consist of a 20\,nm thick epitaxial  SrRuO$_3$ layer (bottom electrode) grown on SrTiO$_3$ (100) by pulsed laser deposition, a 60\,nm Pb(Zr$_{0.2}$Ti$_{0.8}$)O$_3$ thin film and 20\,nm thick Pt top electrodes with a diameter of $d=340$\,$\mu$m. %The voltage applied to the device is measured by an oscilloscope and the current flowing through the device is proportional to the voltage drop on the 50 $\Omega$ load resistor connected to ground.
The experimental design builds on previous studies, in which we investigated the response of similar ferroelectric thin film devices, for which we correlated the structural response - measured by x-ray diffraction \cite{kwamen2017} - with the electrical response for sub-coercive fields\cite{kwamen2019} and high frequencies \cite{Kwamen2023}. In the present study, we added a Q-switched Ti:Sapphire laser system that emits 10\,ns laser pulses at a wavelength of 780 nm. We quantify the fluence $F= 100$\,mJ/cm$^2$ on the sample by measuring the laser spot diameter via a CCD camera at the sample position. %To correlate the laser timing with the electrical pulses, a Si photodiode measures the arrival of the laser pulse via the light scattered at the electrode.

Figure~\ref{fig:Preversal}(a) illustrates the concept of HAPR by showing the free energy $F$ of the system as a function of polarization $P$ modeled as a lor expansion.\cite{rabe2007} Without applied voltage (upper panel) a clear energy barrier separates the two polarization states, even upon considerable heating, as long as $T<T_C$. A strong applied electric field $E$ tilts the potential energy landscape (lower panel), which prefers one ferroelectric polarization over the other. Now the same temperature rise can essentially remove the barrier. Because of the energy barrier, the polarization switching is driven by domain nucleation and growth, according to the Avrami model.\cite{avra1939} 
%Reducing the barrier by transient laser-heating (blue) without electric field may drive the system towards a mixed polarization state. Combining laser heating and an external field (green) may speed up the switching by significant lowering of the barrier and allow for polarization switching. 
Ultimately, suppression of the barrier for few ps and concommitant ultrafast switching should be possible in nanoscale volumes by optimized thermal management.\cite{cahi2014}

%\section{Setup and sample}

Figure~\ref{fig:Preversal}(d) shows a standard hysteresis loop measured via a triangular voltage ramp at 1\,kHz. The remanent polarization $\pm P_\mathrm{r}$ is given by the two values of the polarization for zero applied voltage. Figure~\ref{fig:Preversal}(c) shows the applied voltage sequences, that mimic a classical ferroelectric testing cycle. The black voltage sequence is used to cross check the hysteresis loop from panel (d) via a positive-up-negative down (PUND) sequence. As usual, the-polarization change (black line in panel f) is found as the integral of the measured current. The grey line in panels (b and c) indicates the much shorter first negative pulse, which are used in the laser-assisted measurements. Also the fourth pulse (positive) is shortened to make the excitation symmetric, to  enhance the lifecycle of the device. The pulse sequence loops at 1 kHz synchronized to the 10 ns laser pulse at 780 nm wavelength. Each short electrical field pulse is followed by a saturating long pulse of the same polarity and preceded by a saturating long pulse of opposite polarity. The zoom in Figure~\ref{fig:Preversal}(b) details the timing of the laser pulse $t_\mathrm{L}$ (red) and the voltage pulse duration $t_\mathrm{P2}$ (grey).  

\begin{figure}[thb!]
\centering
\includegraphics[width = 1\columnwidth]{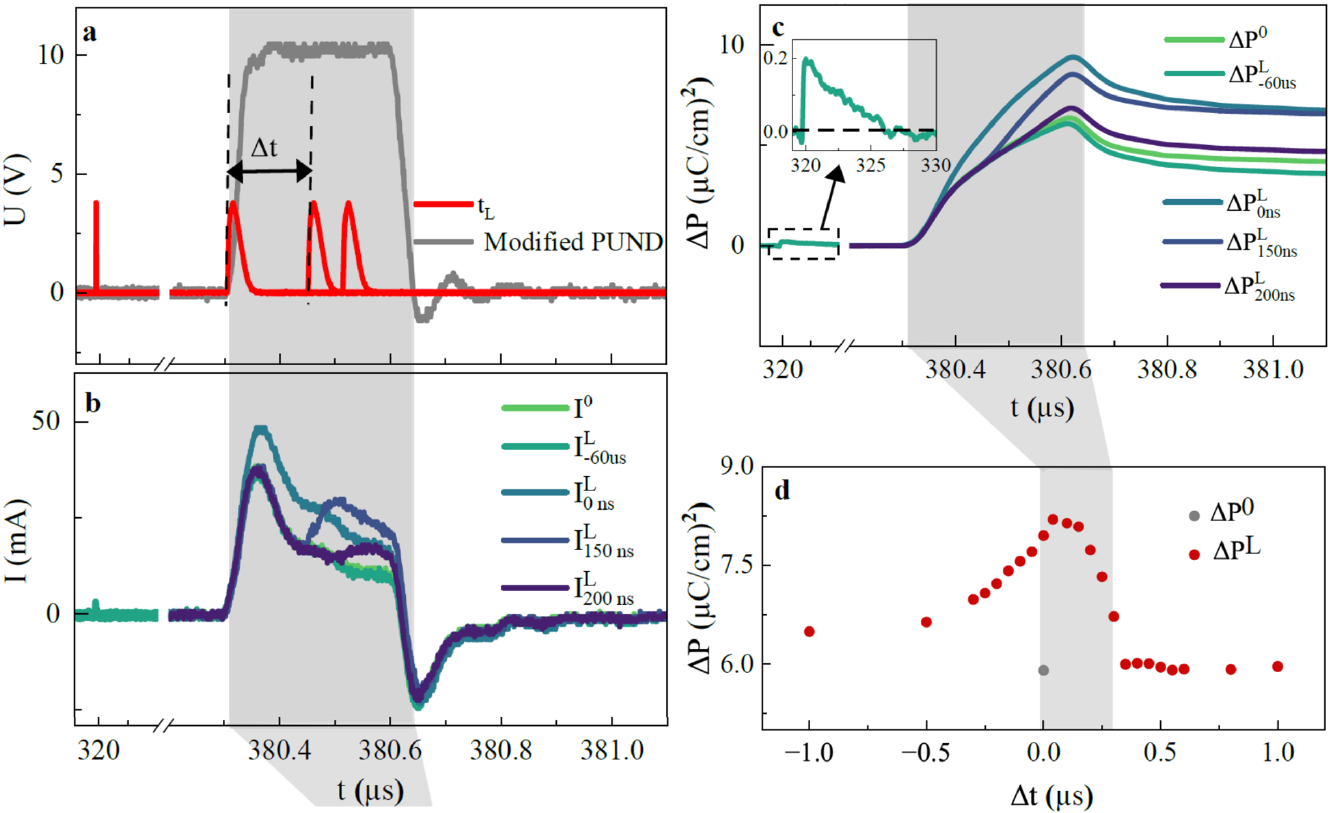}

\caption{(a) Applied voltage pulse (grey) and different laser timing (red) relative to the onset of the voltage pulse. (b) Current response  without laser pulse excitation ($I^0$) and with laser excitation ($I^\mathrm{L}$) for various $t_\mathrm{L}$.  (c) Polarization without laser pulse excitation ($\Delta P^0$) and  with laser excitation ($\Delta P^\mathrm{L}$)  for various $t_\mathrm{L}$. The time window 320–325\,$\mu$s in panels b and c highlights the pyroelectric current and polarization, respectively. The inset in panel c shows the pyroelectric response due to laser excitation when no voltage is applied. (d) Polarization change ($\Delta P$) as a function of laser timing ($\Delta t$) relative to the onset of the voltage pulse. }
\label{fig:delay}
\end{figure}
%\section{Results}
Figure~\ref{fig:Preversal}(e) shows the most relevant result: The grey line indicates the polarization change $\Delta P^0$, which takes place during the 300\,ns negative voltage pulse without laser excitation. The quantitative comparison to the hysteresis loop in Figure~\ref{fig:Preversal}(d) shows that $|\Delta P^0|<P_\mathrm{r}$, i.e. the average polarization is not reversed. The red line, in contrast, indicates that the simultaneous application of the laser pulse increases the polarization change $|\Delta P^\mathrm{L}|>P_\mathrm{r}$ indicating an average polarization reversal of the device. The polarization change is faster under laser-heating and therefore the total amount of switching during the 300 ns electric field pulse is larger. The same result is shown again in Figure~\ref{fig:Preversal}(f) for an extended timescale that allows for a comparison with the result from the standard PUND sequence (black line) that corresponds to the color-coded voltage in panel (c). For technical reasons the data acquisition takes about 1 minute for each trace shown. This corresponds to 60000 laser pulses and the same amount of switching cycles. On the one hand, this demonstrates the robustness of the measurement, on the other hand it is true that in some devices, significant leakage occurs, which increases with measurement time. In order to cross-check that the increased $\Delta P^\mathrm{L}$ in the laser-assisted measurement is indeed due to increased polarization reversal and not due to increased leakage, we additionally quantified the remaining polarization reversal that occurs during the second negative pulse, which has a long duration of 40 $\mathrm{\mu}$s and saturates the polarization. 
Figure~\ref{fig:Preversal}(g) shows that the saturating pulse induces less polarization switching when the laser pulse had been applied during the preceding electrical pulse (red), as partial switching already occurred during the short negative voltage pulse.

\begin{figure}[thb!]
\centering
\includegraphics[width = 0.8\columnwidth]{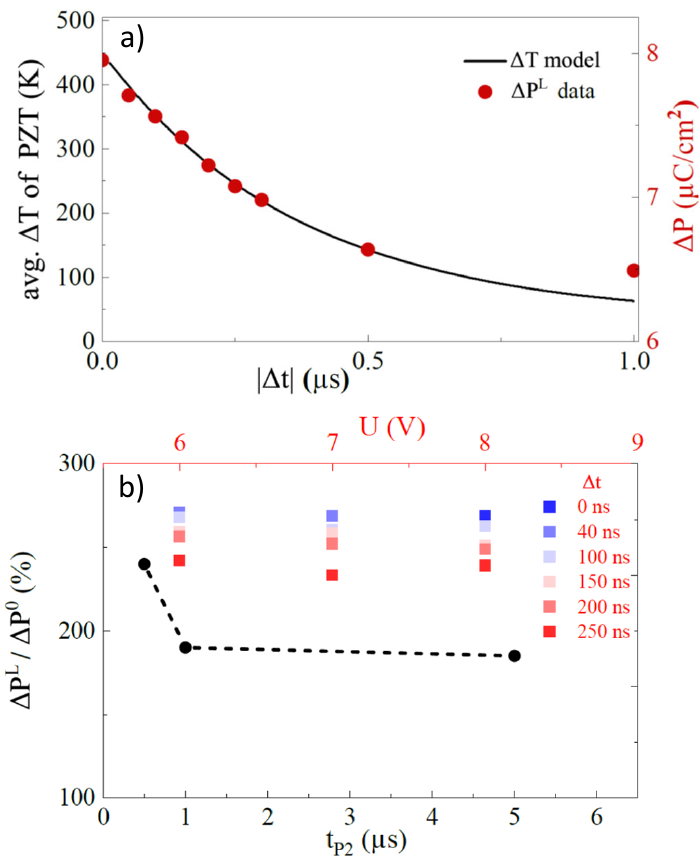}
\caption{a) The average temperature change $\Delta T$ in the PZT film is extracted from modeling the heat transport through the heterostructure. It approximately reproduces the dependence of the laser-induced polarization change $\Delta P^\mathrm{L}$ on the time delay $\Delta t$ between the laser pulse and the electrical pulse. b) dots: Relative polarization enhancement  as a function of voltage pulse duration ($t_{P2}$) with a pulse delay of $\Delta t=0$ and $U=5$\,V. squares: The same for a constant pulse duration $t_{P2}$ = 300 ns  and as a function of applied voltage $U$ for various $\Delta t$. Data in this Figure were recorded on a different device than Fig. 2 and 3}
\label{fig:temperature}
\end{figure}

Fig.~\ref{fig:delay} reports how timing of the laser pulse with respect to the voltage pulses influences the polarization change. Panel (a) shows the short voltage pulse (grey) together with the laser pulse timings (red) for four selected time delays. Note the break of the time axis. The applied voltage  alone induces a current ($I^0$) composed of a positive charging contribution and a positive ferroelectric switching current, as shown in  Panel (b). This ferroelectric contribution additionally increases immediately upon laser excitation, such that the  currents ($I^\mathrm{L}$) depend on the laser arrival time $t_\mathrm{L}$. Upon switching off the voltage, there is a negative current contribution due to the discharging associated with the dielectric contribution of the capacitor. When the laser arrives at  $\Delta t= -60$\,$\mathrm{\mu}$s, i.e. long before the 300\,ns voltage pulse, the response shows a tiny pyroelectric current, in addition to the current induced by the applied voltage pulse only ($I^0$). Panel (c) shows the integrated current signal, i.e. the polarization change without ($\Delta P^0$) and with ($\Delta P^\mathrm{L}$) laser excitation for each timing. For $\Delta t=-60 \mathrm{\mu}$s net zero polarization change due to the purely pyroelectric response is confirmed (seee inset)% highlights the pyroelectric charging which rapidly  peaks upon laser excitation and then decays within $5$\,$\mathrm{\mu}$s as the sample cools. 
The small difference in polarization change   $\Delta P^0$ and %$\Delta P^L_{-60\mathrm{$\mu$}s}$  
$\Delta P^\mathrm{L}$ at $\Delta t=60$\,$\mathrm{\mu}$s can be attributed to  fatigue of this particular device under the current illumination conditions. Although each device shows slightly different fatigue, the presented result are robust. Panel (d) further illustrates the polarization change due to laser excitiation $\Delta P^\mathrm{L}$ measured for  several $\Delta t$ between laser and voltage pulse. $\Delta P^\mathrm{L}$ peaks at $\Delta t=40$\,ns  when the laser pulse coincides with the maximum of the voltage pulse. This suggests that optimal switching  occurs when the thermal activation and the applied voltage act simultaneously. When the laser pulse arrives at $\Delta t> 100$\,ns, $\Delta P^\mathrm{L}$  rapidly falls to the constant $\Delta P^0$  (black dot) which is measured without laser excitation. %This follows from the fact that the laser excitation can only assist the switching at times after its energy has been absorbed. %This persistent change is measured at time delays considerably after the voltage pulse  50 ns apart.  

\begin{table}[b!]
\caption{Relevant thermophysical properties of constituent materials.}
\label{tab:material_properties}
\centering
\begin{tabular}{|c|c|c|c|}
\hline
\textbf{Material} &  \boldmath{$\lambda_\mathrm{pen}$} (nm) & \boldmath{$\kappa_{\text{th}}$ (W/(m$\cdot$K))} & \boldmath{$C_p$ (J/(kg$\cdot$K))} \\
\hline
Pt (20nm) & 7.8 [1] & 5 [2]         & 132 \\
PZT (50nm) & $\infty$    & 5.72 [3]      & 300 [3] \\
SRO (20nm) & $\infty$       & 1 (0.0018) [3] & 481 [3] \\
STO susbtrate & $\infty$       & 10 [3]         & 697 [3] \\
\hline
\end{tabular}

\end{table}

%\begin{figure}[t!]
%\centering
%\includegraphics[width = 0.5 \columnwidth]{fig_5.pdf}
%\caption{dots: Relative polarization enhancement  as a function of voltage pulse duration ($t_{P2}$) with a pulse delay of $\Delta t=0$ and $U=5$\,V. squares: The same for a constant pulse duration $t_{P2}$ = 300 ns  and as a function of applied voltage $U$ for various $\Delta t$. Data in this Figure were recorded on a different device than Fig. 2 and 3}
%\label{fig:Penhancement}
%\end{figure}

When the laser pulse interacts with the electrode at times before the onset of the voltage pulse, i.e., at $\Delta t<0$, the polarization change $\Delta P^\mathrm{L}$ gradually decreases because the ferroelectric has the highest temperature shortly after the laser pulse has been absorbed and cools down as the thermal energy dissipates towards the substrate. In order to check this hypothesis, we implemented a 1D heat transport model in the udkm1Dsim modeling package\cite{schick2021}, assuming literature parameters of the thermophysical constants (see Table 1). We start the modeling with the Pt film at the temperature given by the absorbed photon energy and the rest of the sample at room temperature. Fig.~\ref{fig:temperature}a) shows the calculated temperature change, when we assume a rather poor heat transport across the PZT/SRO interface. The decay of the temperature with time delay $|\Delta t|$ matches very well to the data points that are reproduced from Fig.~\ref{fig:delay}d) for negative $\Delta t<0$. In order to achieve this excellent fit, we assumed a 2\,nm SRO layer with a much poorer heat conductivity of $\kappa=0.0018$\,W/cm$^2$ to model interface resistance. The agreement is taken as an indication that the timescale of heat transport is relevant to the heat assisted switching.

%\section{Discussion}
In Fig.~\ref{fig:temperature}b) we present an additional set of data where laser-excitation approximately doubles the relative amount of polarization change.  This is shown by the black filled circles in the figure for different voltage pulse durations $t_{\mathrm{P2}}$ for a relative delay of $\Delta t=0$ of the laser pulse with respect to the onset of the applied voltage pulse with $U=5\,\mathrm V$. The laser assisted polarization change $\Delta P^\mathrm{L}$ amounts to about 240$\%$ of the switching achieved without laser for $t_{\mathrm{P2}}=0.5$\,ns. For an even shorter fixed pulse duration of $t_\mathrm{P2}= 300$\,ns, we varied the time delay $\Delta t$ (squares) and found the maximum increase of switching efficiency for the smallest $\Delta t =0$\,ns, i.e. when the laser heats the film immediately when the electric field pulse is applied. For voltages ranging from 6 to 8\,V we found increases of switching efficiency up to 275$\%$. When the pulse duration is short, the applied electric field has limited time to induce switching. In this regime, the transient laser heating  significantly lowers the energy barrier, strongly enhancing the polarization reversal. For longer pulses, some of the laser-induced heat dissipates before the switching process completes for a given duration, reducing polarization reversal. 

%The device that has been used to acquire the data shown in Fig.~5 unfortunately started to deteriorate, before we could reproduce the absolute calibration of the remanent polarization after the measurement. Since the fatigue decreases the total amount of switchable polarization, the effect may be even stronger, however, we could not ascertain that $\Delta P^\mathrm{L}>P_\textrm{r}$. 
We are convinced that much better contrast ratios can be achieved with optimized devices and optimized voltage and laser pulse sequences. In particular, smaller area and improved electrical contact, as well as thinner electrodes and ferroelectric films, should drastically decrease the heat load and concomitant fatigue. The voltage pulse duration is ideally as short as the timescale over which heating is significant. In smaller structures this can easily go down to 100\,ps, and then also similarly short voltage pulses can be used, which significantly decreases the polarization switching without laser pulse. Different material compositions and strain conditions may help to improve the domain wall propagation velocity. The faster the process and the smaller the ferroelectric volume that needs to be heated, the less energy will be consumed for the laser-assisted switching. 
%As an estimate we take a 10 nm thick BaTiO$_3$ film with a 200 nm electrode diameter to calculate an energy content of the light pulse of about $W=C_V V \Delta T = 0.1$ pJ per pulse, using a specific heat of $2.6\cdot10^6$J/($m^3$K) and a temperature increase of $\Delta T = 100$K.
As an order of magnitude estimate for the small energy of the light pulse required to raise the temperature by $\Delta T = 100$\,K for an optimized barrier, we take a 10\,nm thick ferroelectric film with a 200 nm electrode diameter. The pulse energy is only $W=C_\mathrm{V} V \Delta T = 0.1$\,pJ, using a specific heat of about $2.6\cdot10^6$J/(m$^3$K). %Thus, the switching can become rather energy efficient.% compared to an average value of 1\,nJ per bit.  

%\section{Conclusion}
In conclusion, our experimental results show that ferroelectric polarization switching is enhanced when a laser pulse deposits thermal energy into the metallic electrode, which is subsequently transferred to the adjacent ferroelectric thin film, thereby lowering the energy barrier.
The enhancement is robust across various applied voltages and time delays $\Delta t$ between laser and voltage pulses. Under favorable conditions, the combination of heat and electric field can lead to switching of the average polarization direction. This laser-assisted polarization switching is indeed a heat-assisted polarization reversal (HAPR), since the light only interacts with the metal electrode, which increases the temperature of the ferroelectric. Optimization of the device material and thickness together with proper lateral size scaling and heat management should improve the switching contrast dramatically. These advancements make HAPR a promising approach for the development of energy-efficient electronics, non-volatile memory devices, and next-generation data storage technologies. 

\section*{References}
\bibliography{bib.bib}

%\subsection{Acknowledgements}
We acknowledge funding by the Deutsche Forschungsgemeinschaft (DFG) and French Agence Nationale de la Recherche (ANR) via the FEAT-project DFG 431399790/ANR-19-CE24-0027-01, and the BMBF for funding project No. 05K22IP1. We thank Alan Brunier from the University of Warwick for preparing the ferroelectric devices.
\subsection{Data availability}
Data recorded for the experiment are available from the authors upon reasonable request.

%\subsection{Author contributions}
%JEP, AM, MB supervised and coordinated the project
%JEP, MH, conceived and conducted Experiment.
%ARF, RS, UB prepared and setup the experiment
%JEP, UB, ARF, RS, WJ, WL, FB operated the MID instrument
%MiH created the sample
%MH, MM, AvR characterized the sample (x-ray, optical)
%JEP, MM, DS conducted the simulations
%MH, MB added theoretical background for SL
%JEP performed the data evaluation with help of JW, JM
%All authors contributed to the manuscript.

\end{document}